\documentclass[%
 aip,jcp,
amsmath,amssymb,
reprint,
]{revtex4-1}

\usepackage{slantsc}
\usepackage{lmodern}
\usepackage{graphicx}% Include figure files
\usepackage{dcolumn}% Align table columns on decimal point
\usepackage{amsmath}
\usepackage{bm}        % for math
\usepackage{amssymb}   % for math
\usepackage{xcolor}

\begin{document}
\preprint{AIP/123-QED}

\title{Optimal \textit{in situ} electromechanical sensing of molecular species}% Force line breaks with \\
%\thanks{Footnote to title of article.}

\author{Maicol A. Ochoa}
\affiliation{Biophysics Group, Microsystems and Nanotechnology Division, Physical Measurement Laboratory, National Institute of Standards and Technology, Gaithersburg, MD 20899}
\affiliation{Maryland Nanocenter, University of Maryland, College
Park, MD 20742}

\author{Michael Zwolak}
\email{mpz@nist.gov}
\affiliation{Biophysics Group, Microsystems and Nanotechnology Division, Physical Measurement Laboratory, National Institute of Standards and Technology, Gaithersburg, MD 20899}
 %\homepage{http://www.Second.institution.edu/~Charlie.Author.}
%\affiliation{%

%Second institution and/or address%\\This line break forced% with \\
%}

\date{\today}% It is always \today, today,

\begin{abstract}
  We investigate protocols for optimal molecular detection with electromechanical nanoscale sensors in ambient conditions. Our models are representative of suspended graphene nanoribbons, which due to their piezoelectric and electronic properties, provide responsive and versatile sensors. In particular, we analytically account for the corrections in the electronic transmission function and signal-to-noise ratio originating in environmental perturbations, such as thermal fluctuations and solvation effects. We also investigate the role of the sampling time in the current statistics. As a result, we formulate a protocol for optimal sensing based on the modulation of the Fermi level at fixed bias, and provide approximate forms for the current, linear susceptibility, and current fluctuations. We show how the algebraic tails in the thermally broadened transmission function affect the behavior of the signal-to-noise ratio and optimal sensing. These results provide further insights into the operation of graphene deflectometers and other techniques for electromechanical sensing.
%\verb+\pacs{#1}+ command.
\end{abstract}

%Use showkeys class option if keyword
\maketitle

\section{Introduction}

\vspace{-0.4cm}
Nanoscale devices that integrate two-dimensional piezoelectric materials -- such as graphene nanoribbons (GNR) -- are feasible alternatives for electromechanical molecular detection\cite{wu2004computational, bell2006three, stampfer2006nano, obitayo2012review,cullinan2012scaling,lyshevski2018nano, hu2019nonlinear,su2019electromechanical} at room temperature, and in complex environments. This in turn will provide new venues for electronic-based biomolecular analysis\cite{storm2005fast,lagerqvist2006fast,smeets2006salt,lagerqvist2007comment,lagerqvist2007influence,zwolak2008colloquium,liu2010translocation,chang2010electronic,schneider2010dna,huang2010identifying,heerema2016graphene,di2016decoding,heerema2018probing}. In this setting, molecular sensing is possible as a result of the modifications in the transport properties of the GNR due to interactions with the analyte, and is limited by noise originating in environmental fluctuations. The latter may significantly alter the performance of the device at room temperature and in wet ionic solutions. Indeed, we recently showed that the electronic transmission function turns into a generalized Voigt profile under the influence of inhomogeneous conditions\cite{ochoa2019generalized}. Numerical investigations in graphene deflectometry\cite{gruss2017communication, gruss2018graphene} -- a proposed detection technique for single molecules that correlates the local deflection of the graphene nanoribbon with the current -- illustrate that thermally-induced mechanical fluctuations increase the noise and affect the conditions for optimal detection.

The electronic conductance through nanoconfined systems, such as in molecular break junctions\cite{frisenda2013statistical,kim2014determination}, is influenced by the local structure, variations of which are sampled during repeated formation of the junction. This variation is fitted to Gaussian\cite{kim2014determination} and other distributions. In this case, the complexity in the histograms is of structural origin, such as the heterogeneity in the orientation of the molecule at the junction. The histogram of currents takes a particular form in the presence of external mechanical forces\cite{franco2011tunneling,koch2018structural,mejia2019force}, with the structural fluctuations affecting both force and conductance.  In the off-resonance regime and under elastic conditions, the conductance histogram is an indirect map of the values taken by the transmission function.

In this paper, we investigate optimal protocols for electromechanical sensing at room temperature. We demonstrate that one must account for the environmental effects imprinted in the Voigt (or generalized Voigt) lineshape of the transmission function to design optimal sensing protocols. To see this, we consider three cases: a fully Gaussian picture, exact numerical solution, and an approximate form for the Voigt distribution.  Under ambient conditions, the current distribution can take complex forms that vary with the experimental sampling time. For large sampling times, every distribution approaches to its Gaussian limit, by the law of large numbers. We identify for a representative Langevin model the shortest sampling time, in terms of friction coefficient and reduced mass, for which fluctuations decrease as the inverse of the square root of sampling. When the bias window -- the difference in the electron distribution between the contacts -- is larger than the energy fluctuations induced by environment, we provide accurate analytical estimates for the current, fluctuations, and SNR (in the absence of noise due to the readout electronics) for electromechanical sensing, and demonstrate that it is necessary to account for the algebraic tails in the Voigt transmission function for fast sampling times and far away from the molecular level.

\vspace{-0.05cm}
 The organization of the paper is as follows. In section \ref{sec:gauss}, we consider an approximate Gaussian fit for the transmission function of a single level and investigate optimal detection in a protocol that modulates the Fermi level at fixed bias. This gives a fully analytical -- albeit approximate -- approach to understanding some of the basic aspects of the optimization problem. Next, in Sec.~\ref{sec:time}, we analyze the effect of the sampling time in the current distribution and show that for large enough sampling times, the current distribution converges to a Gaussian form. Finally, in Sec.~\ref{sec:lorentzian}, we revisit the optimal protocol for sensing accounting for the full Voigt profile in the Gaussian limit for the current distribution. We summarize in Sec.~\ref{sec:conclusion}.

\vspace{-0.5cm}
%%%%Equation -- text spacing%%%%
%\setlength{\belowdisplayskip}{3pt} \setlength{\belowdisplayshortskip}{3pt}
%\setlength{\abovedisplayskip}{3pt} \setlength{\abovedisplayshortskip}{3pt}

 \section{Gaussian Model}\label{sec:gauss}
 \vspace{-0.4cm}
 Transport properties in the characteristic regimes of molecular structures can be analyzed by representing the system with a single level\cite{datta1997electronic, gruter2005resonant,huisman2009interpretation, zotti2010revealing}, even when the distribution of the transmission function deviates from a simple Lorentzian\cite{kim2014determination, quan2015quantitative}. This approach can be used to investigate non-interacting tight-binding models, which in the case of graphene itself, are accurate\cite{wallace1947band, zheng2013tight, nakada1996edge,hancock2010generalized}. For a single level, we start with the fully Gaussian problem due to its tractability and to illustrate some of the expected general principles. As an implication of the central limit theorem, current distributions should converge to a Gaussian form for long sampling times. Thus, a Gaussian model should be representative, with corrections dependent on the sampling time and also the Gaussian approximation to the bias window (discussed in the Supplementary Material). We consider a Gaussian fit to the transmission function $T$ of a single level $\varepsilon_p$ with homogeneous broadening $\sigma_T$
 \begin{align}\label{eq:gaussT}
 T ={}& A e^{-\frac{(\varepsilon -\varepsilon_p)^2}{2 \sigma_T^2}},
\end{align}
and investigate modifications in $T$ resulting from the {\it inhomogeneous} fluctuations in the level energy $\varepsilon_p$ due to the noisy environment. In Eq.~\eqref{eq:gaussT}, $A$ is a normalization constant independent of the level energy. We assume that the source of inhomogeneous broadening modifies $\varepsilon_p$ around its equilibrium value $\bar \varepsilon_p$ according to the Gaussian distribution
\begin{align}
  g(\varepsilon_p - \bar \varepsilon_p) ={}&  \frac{1}{\sqrt{2 \pi \sigma_{\rm S}^2}} e^{- \frac{(\varepsilon_p - \bar \varepsilon_p)^2}{2 \sigma_{\rm S}^2}}.\label{eq:g_e}
\end{align}
For some systems\cite{gruss2018graphene,ochoa2019generalized}, $\sigma_{\rm S}^2$ is proportional to the environmental temperature. In particular, for sensors made of graphene nanoribbons, the most important fluctuations are on the order of nanoseconds and are well-separated from the timescale for electron transport. Moreover, we assume that the length of the suspended structure is shorter than the mean-free-path for an electron in graphene. The average thermally broadened form of the transmission function is 

\begin{align}
  \langle T \rangle ={}& \int d\varepsilon_p T(\varepsilon - \varepsilon_p) g(\varepsilon_p - \bar \varepsilon_p)\\
 ={}&  \frac{A\, \sigma_T}{\sqrt{\sigma_T^2 + \sigma_{\rm S}^2}} e^{- \frac{(\varepsilon - \bar \varepsilon_p)^2}{2 (\sigma_T^2 +\sigma_{\rm S}^2)}}.\label{eq:T_broad_gauss}
\end{align}
 As a result, thermal broadening of a Gaussian transmission function does not change the qualitative form of the transmission function, but modifies its spread as the contribution of two independent mechanisms. The stationary current is given by the Landauer-B\"uttiker formula 
\begin{align}\label{eq:Iavg}
  \langle I \rangle  ={}& \frac{2 e}{\hbar} \int \frac{d \varepsilon}{2 \pi}  \langle T (\varepsilon - \bar \varepsilon_p) \rangle \left[ f_{\mathcal{L}}(\varepsilon) - f_{\mathcal{R}}(\varepsilon) \right],
\end{align}
in terms of $\langle T \rangle$, and where $f$ denotes the Fermi function $f_{\mathcal{L}/\mathcal{R}}(\varepsilon) =~[\exp(\beta (\varepsilon - ~\mu_{\mathcal{L}/\mathcal{R}}) + 1]^{-1}$, $\mu_{\mathcal{L}/\mathcal{R}}$ is the chemical potential at the left/right contact and $\beta$ is the inverse temperature.
This form of the current depends on the separation of the timescales between electronic and atomic dynamics, and the fact that environmental fluctuations ensure that atomic coherences are rapidly suppressed. In the discussion below, we consider that a symmetric bias of magnitude $\Delta \mu$ is applied to the system, such that $\mu_{\mathcal{L}} = \mu + \Delta \mu/2$ and $\mu_{\mathcal{R}} = \mu - \Delta \mu/2$, with Fermi energy $\mu$. At room temperature and under small bias, an accurate approximation to the bias window (BW) is given by the form   
\begin{align}\label{eq:Bias}
f_{\mathcal{L}}(\varepsilon) -f_{\mathcal{R}} (\varepsilon) \approx{}& \tanh \left( \frac{\beta \Delta \mu}{4}\right) e^{-\frac{(\varepsilon-\mu)^2}{2 \sigma_{\rm BW}^2}},
\end{align}
where $\sigma_{\rm BW}^2$ is a measure of the bias window broadening determined by the full-width at half maximum $\sigma_{\rm BW} \beta \sqrt{2 \ln 2} = {\rm arccosh}(2 + \cosh(\Delta \mu \beta/ 4))$. From Eqs.~\eqref{eq:T_broad_gauss},~\eqref{eq:Iavg}, and \eqref{eq:Bias} we obtain a closed form for the inhomogeneous average of the current $\langle I \rangle$
\begin{align}
  \langle I \rangle  ={}& I_{\rm G}  e^{-\frac{(\bar \varepsilon_p -\mu)^2}{2 \sigma^2}},\label{eq:Igauss}
\end{align}
where $I_{\rm G}$ is given by
\begin{align}
  I_{\rm G} ={}& \frac{2 e}{ h} \frac{A \sigma_T \sigma_{\rm BW}}{\sqrt{2 \pi} \sigma} \tanh \left(\frac{\beta \Delta \mu}{4} \right),
\end{align}
and $ \sigma^2 = \sigma_T^2 + \sigma_{\rm S}^2 + \sigma_{\rm BW}^2$.  A closed form for the thermally-broadened linear susceptibility $\chi_{\varepsilon}$ also follows from Eq.~\eqref{eq:Bias} 
\begin{align}
 \chi_\varepsilon ={}& \frac{2 e}{\hbar} \int \frac{d \varepsilon}{2 \pi}  \langle \partial_{\bar \varepsilon_p} T \rangle \left[ f_{\mathcal{L}}(\varepsilon) - f_{\mathcal{R}}(\varepsilon) \right],\\
  ={}&\frac{1}{\sigma^2} (\bar \varepsilon_p - \mu) \langle I \rangle \label{eq:chiG},
\end{align}
which indicates that for a Gaussian fit to the transmission function the linear response is proportional to the average current.

Next we account for fluctuations in the current $\langle I \rangle$ originating in the inhomogeneous environment, and consider the variance in the current distribution $\sigma_I^2 = \langle I^2 \rangle  -\langle I \rangle ^2$. For a given realization of the energy level $\varepsilon_p$, the instant current through the level is
\begin{align}
     I(\varepsilon_p) ={}& \frac{2 e}{\hbar} \int \frac{d \varepsilon}{2 \pi}  T (\varepsilon - \varepsilon_p)  \left[ f_{\mathcal{L}}(\varepsilon) -f_{\mathcal{R}} (\varepsilon) \right],\label{eq:Iinstant}\\
  ={}& \frac{I_{\rm G}\, \sigma}{\sqrt{\sigma_T^2 + \sigma_{\rm BW}^2}} e^{- \frac{(\varepsilon_p -\mu)^2}{2 (\sigma_T^2 + \sigma_{\rm BW}^2)}}.\label{eq:Io_gauss}
\end{align}
This allows us to compute $\langle I^2 \rangle$ and $\sigma_I^2$, and obtain
\begin{align}
  \langle I^2 \rangle ={}& \int d\varepsilon_p I(\varepsilon_p )^2 g(\varepsilon_p - \bar \varepsilon_p)\\
  ={}& \frac{I_{\rm G}^2 \sigma^2}{ \sqrt{(\sigma^2 - \sigma_{\rm S}^2)(\sigma^2 + \sigma_{\rm S}^2)}} e^{-\frac{(\bar \varepsilon_p - \mu)^2}{ \sigma^2 + \sigma_{\rm S}^2}}
\intertext{and}
  \sigma_I^2 ={}& I_{\rm G}^2 \,\sigma^2 \left(\frac{e^{-\frac{(\bar \varepsilon_p - \mu)^2}{ \sigma^2 + \sigma_{\rm S}^2}}}{\sqrt{\sigma^4 - \sigma_{\rm S}^4}} - \frac{e^{-\frac{(\bar \varepsilon_p -\mu)^2}{\sigma^2}}}{\sigma^2} \right)\label{eq:sigmaIG}.
\end{align}

\noindent This is always positive\footnote{ To show this note that $\sqrt{\sigma^4 - \sigma_{\rm S}^4} \leq \sigma^2$, and consider that $e^{-z^2}$ is a decreasing function on $z$} and, in the absence of thermal fluctuations ($\sigma_{\rm S} =0$), it vanishes. The quantity $\sigma_I$ captures the excess fluctuations in the current induced by the local environment.

 A protocol for electromechanical detection that records the current -- or more precisely changes in the current -- at a fixed bias $\Delta \mu$, can be optimized in terms of the Fermi level $\mu$.  The results in Eqs.~\eqref{eq:chiG} and \eqref{eq:sigmaIG} provide the following analytical estimates for optimal detection.

 For a given shift in the equilibrium energy level $\Delta \varepsilon~=~\varepsilon_p ~- \bar \varepsilon_p$, the maximal change in the current magnitude, $|\Delta \langle I \rangle| = |\langle I (\bar \varepsilon_p)\rangle-\langle I (\bar \varepsilon_p+\Delta \varepsilon_p)\rangle|$, is obtained from $\left| \partial_{\mu} \chi_\varepsilon \right|= 0$. In terms of the Fermi level $\mu$, this maximum is achieved at
\begin{align}
  \mu^{*}_{\Delta I} = \bar \varepsilon_p \pm \sigma .\label{eq:mu_DI}
\end{align}
This result is natural for the Gaussian form for $\langle I \rangle$ found in Eq.~\eqref{eq:Igauss}. The maximum change in the current for a small change in peak position ($\bar \varepsilon_p$) occurs when the derivative with respect to $\bar \varepsilon_p$ is maximal. This happens at $\pm \sigma$ from the peak for a Gaussian.

  Current fluctuations, as accounted for by $\sigma_I$, have a local minimum when $\mu = \bar \varepsilon_p$, and are maximal at\footnote{This value can be found by solving $\partial _{\mu} \sigma_I^2 =0$ in terms of $\mu$. } 
  \begin{align}
    \mu^{*}_{\sigma_I} ={}& \bar \varepsilon_p  \pm\frac{\sigma}{\sigma_{\rm S}}\sqrt{(\sigma^2+\sigma_{\rm S}^2)} \times\notag\\
    &\hspace{0.7cm} \sqrt{\ln \left[\frac{(\sigma^2+\sigma_{\rm S}^2)\sqrt{\sigma^4 - \sigma_{\rm S}^4}}{\sigma^4}\right]},\label{eq:mu_sigmaI}
  \end{align}
  when the ratio $\sigma_{\rm S}^2/ \sigma^2$ is between zero and approximately\footnote{Note that $\mu_{\sigma_I}^{*}$ in Eq.~\eqref{eq:mu_sigmaI} is a real number when the argument in the logarithm is greater than one. Letting $\gamma = \sigma_{\rm S}^2/\sigma^2$, we rewrite the argument in terms of $\gamma$ 
    \begin{align*}
      (1+\gamma)^2(1-\gamma^2) =1,
    \end{align*}
    which has as real solution $\gamma = 0$ and
    \begin{align}
      \gamma = \frac{1}{3}\left[\left(19 - 3 \sqrt{33} \right)^{1/3}+\left(19 + 3 \sqrt{33} \right)^{1/3}-2\right]\approx 0.839
    \end{align} 
  } $ 0.839$.  Beyond this range, thermal fluctuations dominate and $\sigma_I$ is maximal at the current maximum ($\mu = \bar \varepsilon_p$). On the other hand, the result in Eq.~\eqref{eq:mu_sigmaI} simplifies substantially when the bias window is large compared to the thermal fluctuations (i.e., $\sigma_{\rm S} < \sigma_{\rm BW}$)
  \begin{align}
  \mu^{*}_{\sigma_I}(\sigma_{\rm S} < \sigma_{\rm BW}) \approx {}& \bar \varepsilon_p  \pm \sigma \left(1 - \frac{\sigma_{\rm S}^4}{2 \sigma^4} \right),
  \end{align}
  and in the limit of weak thermal fluctuations $\sigma_{\rm S} \ll \sigma $, $\mu^{*}_{\sigma_I}$ and $\mu^{*}_{\Delta I}$ coincide. 

  An optimal protocol for sensing must maximize the signal-to-noise ratio SNR defined by
\begin{align}
  {\rm SNR} = \frac{|\Delta I|}{\sigma_I} =\frac{|\chi_\varepsilon \Delta \varepsilon|}{\sigma_I}, 
\end{align}
For the proposed sensing protocol and when the bias window is large (i.e., $\sigma_{\rm S} < \sigma_{\rm BW}$), optimal values are approximately achieved at\footnote{This result is the solution set for $\partial_{\mu} {\rm SNR} = 0$, and utilizes the following identity
\begin{align}
  \frac{\partial}{\partial \mu} \left( \frac{\Delta I}{\sigma_I} \right) ={}& \frac{\Delta \varepsilon}{\sigma_I^3}\left[\partial_\mu \chi_\varepsilon \sigma_I^2 - \chi_\varepsilon  \sigma_I \partial_\mu \sigma_I \right].
\end{align}
}
\begin{align}
  \mu^{*}_{\rm SNR} \approx {}& \, \bar \varepsilon_p \pm \frac{\sigma}{\sigma_{\rm S}}\sqrt{(\sigma^2 + \sigma_{\rm S}^2) \ln \left[ \frac{\sigma}{\sqrt{\sigma^2 - \sigma_{\rm S}^2}} \right]}. \label{eq:MaxSNRGauss} 
\end{align}

\noindent Again, we can provide a simpler form in the case of a large bias window
\begin{align}
  \mu^{*}_{\rm SNR} \approx {}& \bar \varepsilon_p \pm \frac{\sigma}{\sqrt{2}} \left(1+\frac{3}{2}\frac{\sigma_{\rm S}^2}{\sigma^2}+\frac{1}{4}\frac{\sigma_{\rm S}^4}{\sigma^4}\right). \label{eq:MaxSNRGausslim} 
\end{align}

Notice that when thermal fluctuations are small (i.e., $\sigma_{\rm S} \ll \sigma$), the maximum in the SNR occurs at  $\mu^{*}_{\rm SNR}=~\bar \varepsilon_p \pm  \sigma/\sqrt{2}$, that is, closer to the current maximum than the maximal response $\mu^{*}_{\Delta I}$ in Eq.~\eqref{eq:mu_DI}. We note that some of these considerations depend on the bias window approximation. However, the Gaussian bias window approximation works well near the Fermi level (i.e., within $\pm \sigma$) and thus the expressions are accurate for the important cases.
\vspace{-0.3cm}

\section{Sampling time and normal distribution}\label{sec:time}
\vspace{-0.1cm}

The statistical properties in the current are also determined by the sampling time $\tau$. Different mechanisms of electronic and structural relaxation, and electron transfer (intramolecular and to the contacts) contribute at different timescales to the total noise and broadening. For large enough sampling times, structurally induced fluctuations in the current naturally converge to a Gaussian distribution. 
 \noindent We emphasize that the sources of randomness in the current that we investigate here originate in thermal and environmental fluctuations, and are different from those due to geometric factors, such as device-to-device structural variations and randomness in the binding strength to the contacts. The latter histograms treat the structure as static, in which mechanical fluctuations are averaged over during a current measurement.
 
 We start by considering a simple model, in which the energy of the level varies as a function of a structural parameter $Y$, which we assume follows a Langevin equation of motion.  We are interested in sensing protocols at room temperature and complex environments where the atomic dynamics dephase rapidly. Moreover, for short nanoribbons, an electron injected at the Fermi energy should cross in the 10's of femtoseconds, allowing for multiple reflection at the electrode interfaces (the transit across one length of the nanoribbon is even less). These conditions are sufficient for a classical description of the atomic motion and fluctuations, while electron dynamics are calculated from quantum mechanical principles. Thus, energy oscillations originate on random forces acting on $Y$, subject to a relaxation process with characteristic friction coefficient $\eta$ and spring constant $\kappa$. For this model, fluctuations in the current are determined by those in $Y$ around the equilibrium value $\bar Y$ (i.e., $\delta Y = Y - \bar Y$). To first order in this parameter $\sigma_I^2 \approx \sigma_Y^2 \left(\partial_Y I (\bar Y)\right)^2$.  We obtain $\sigma_Y^2$ from the time correlation function for the parameter $Y$ (see Appendix~\ref{ap:x_variance}),  and find that 
\begin{align}
  \sigma_I^2 \propto \sigma_Y^2 ={}& \frac{\eta}{\tau \kappa^2 \beta}, \hspace{0.3cm} (\text{first order})
\end{align}
for sampling times $\tau > \eta^{-1} m $, where $m$ is the mass of the oscillator. Next, we investigate the linear mechanical susceptibility $\chi_Y$. By considering linear deviations from equilibrium interatomic distances $\Delta Y$ in Eq.~\eqref{eq:Iavg}, we obtain the linear response in the stationary current in the form $\Delta I = \chi_Y  \Delta Y$, and we notice that $\chi_Y$ is independent of the sampling time\footnote{Notice that $\bar I$ is independent of sampling time $\tau$, and so is the difference $\Delta \bar I$ induced by the shift $\Delta Y$.}. Consequently, we also show that
\begin{equation}
  \label{eq:SNR_1st_order}
  {\rm SNR } \propto \sqrt{\frac{\kappa^2 \beta \tau}{\eta}}. 
\end{equation}
Thus, as usual, the SNR improves as one increases the sampling time as the square root of the sampling time and deteriorates when one increases the temperature or the mechanical friction. This is a standard result for sampling, which is relevant to deflectometry in hot, wet environments.
  
%%%%%%%%%%%%%%%%%%%%%Fig1%%%%%%%%%%%%%%%%%%%%%%%%
\begin{center}
  \begin{figure}[t]
    \centering
    \includegraphics[scale=0.28]{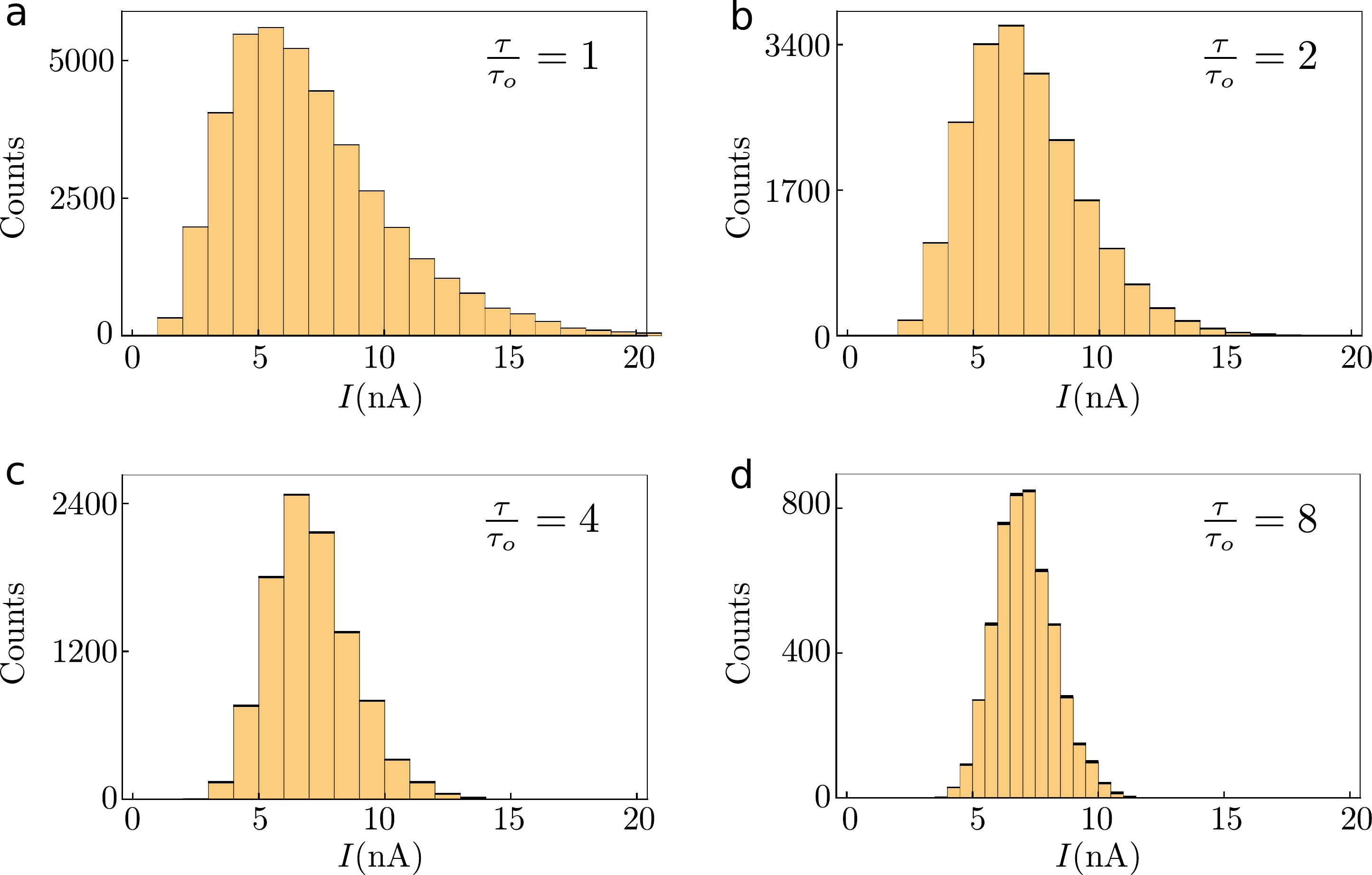} 
    \caption{(Color online) Current histograms for different sampling times $\tau  = n \tau_o$, where  $\tau_o$ is the minimum time for which two sequential readings in the current are independent. A random sequence in the current is generated by sampling $\{ \varepsilon_p^{(i)}\}$ according to Eq.~\eqref{eq:g_e}, and for each value $\varepsilon_p^{(i)}$ we compute the current $I(\varepsilon_p^{(i)})$, as in Eq.~\eqref{eq:Iinstant}. (a) $n = 1$ (b) $n = 2$ (c) $n = 4$ (d) $n = 8$. Parameters for this model are taken from Ref.~\citenum{ochoa2019generalized} and are chosen to reproduce the first peak in the transmission function for a suspended graphene nanoribbon: $\varepsilon_p = 0.153$~eV, $w = 1.3$~meV, $\sigma_{\rm S} = 14$~meV (such that, $\sigma_{\rm S}^2 = 7.57~\text{meV}/\beta$),  $\mu = 0.07$~eV, $\Delta \mu = 50$~meV and $300$~K.  These results are for an initial sample size of 40000, representing measurements at $\tau= \tau_o$, and that we rescale proportionally for larger sampling times.  Thus, the number of points used in each histogram is (a) 40000, (b) 20000, (c) 10000, and (d) 5000. The standard deviation for each bin is less than 6 counts, and it is smaller than the thickness of bin borderline.}
    \label{fig:histogram}
  \end{figure}
\end{center}
%%%%%%%%%%%%%%%%%%%%%%%%%%%%%%%%%%%%%%%%%%%%%%%%%   

\vspace{-1.0cm}
The above discussion followed from the observation that for large sampling times, as compared to internal relaxation processes, the correlation between sequential events (i.e., the memory of the system) diminishes (Appendix \ref{ap:x_variance}). More generally, we can consider that the active material in the electromechanical sensor system has a characteristic time $\tau_o$, for which two sequential readings in the current are independent. For a suspended graphene nanoribbon, this time is given by the relaxation time to a new independent configuration. In other words, the current read over a timescale $\tau_o$ gives one independent sample from the energy space for $\varepsilon_p$. Measuring at time $\tau > \tau_o$ must therefore provide $\tau/\tau_o = n$ independent reads\footnote{By measuring the current over shorter times, $\tau < \tau_o$, we are most likely to sample over one configuration. In this case, the main source of randomness is the shot noise. In the limit of weak coupling to the contacts and in the absence of structural randomness, shot noise follows a Poisson distribution (see Ref.~\citenum{blanter2000shot}).}. For a family of $n$ current readings, let us define a new random variable $I_\tau~=~\frac{1}{n}\sum_{i=1}^n I_i$, corresponding to the measured current for a sampling time $\tau = \tau_o n$. Then the first three central moments for the distribution of $I_\tau$ are (Appendix \ref{ap:skewness})
\begin{align}
  \langle I_\tau - \bar I \rangle_o ={}& 0,\\
  \langle (I_\tau - \bar I )^2 \rangle_o ={}& \frac{\tau_o}{\tau} \sigma_I^2,\label{eq:sigma2}\\
  \langle (I_\tau - \bar I )^3\rangle_o ={}& \left(\frac{\tau_o}{\tau}\right)^2 s_3 ,\label{eq:skewN}
\end{align}
where $s_3$ is the third moment (skewness) for the current distribution, and $\langle \rangle_o$ is the arithmetic mean. In the usual way, the variance decreases as the inverse of the number of independent measurements  while the expectation value does not change.

\begin{center}
  \begin{figure}[t]
    \centering
    \includegraphics[scale=0.5]{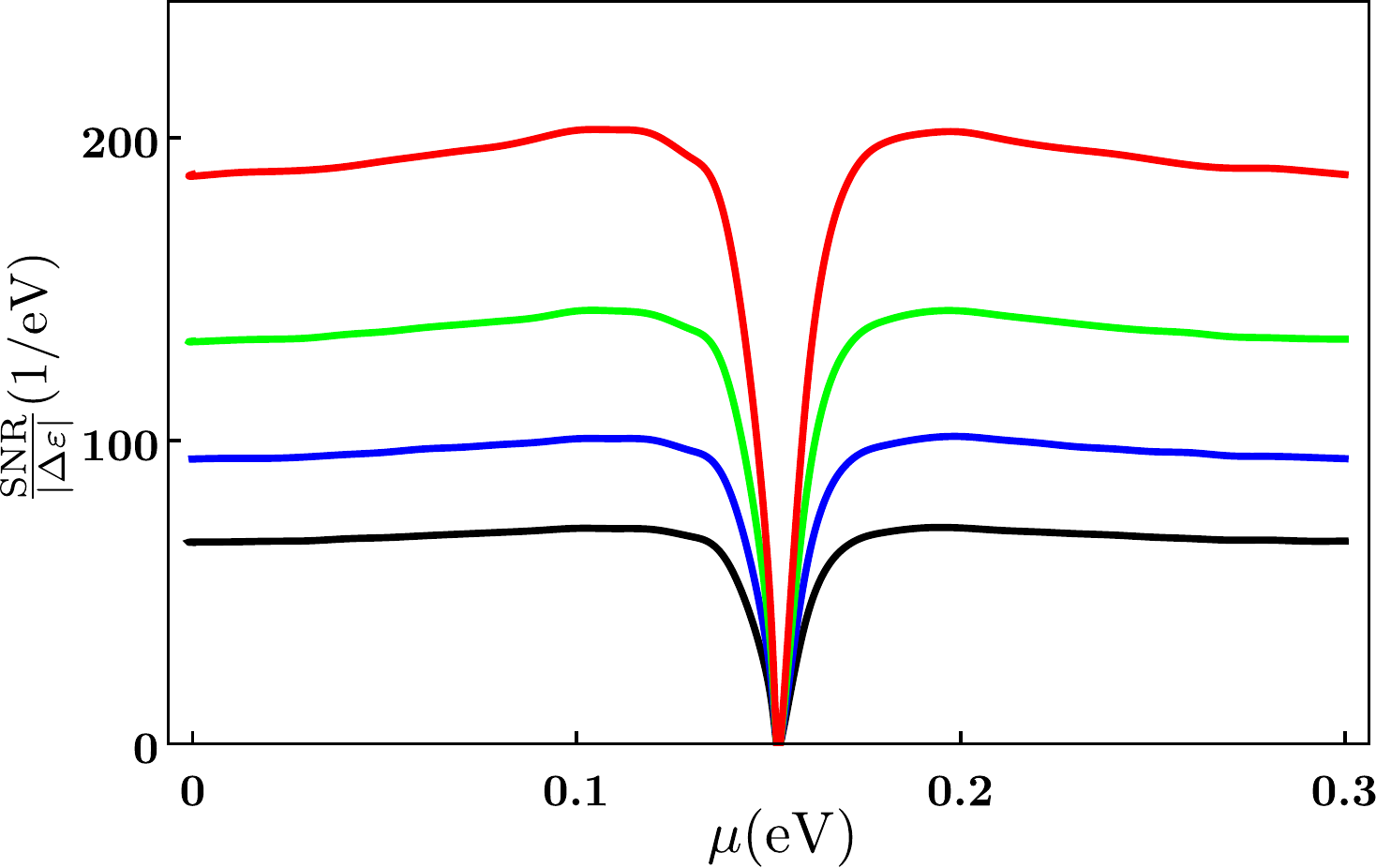} 
    \caption{(Color online) SNR as a function of the Fermi energy $\mu$ for different sampling times $\tau = n \tau_o$.  (black) $n = 1$ (blue) $n = 2$ (green) $n = 4$ (red) $n = 8$. Parameters for this model are the same than those in Fig. \ref{fig:histogram}.}
    \label{fig:SNR_sampling}
  \end{figure}
\end{center}

  Another important remark resulting from Eq.~\eqref{eq:skewN} is that the current distribution should quickly converge to a Gaussian distribution. In Fig.~\ref{fig:histogram} we numerically observe this convergence by following the evolution of the histograms for the current for several sampling time ratios $n$. Significantly, Fig.~\ref{fig:histogram} shows that at room temperature the current distribution for our model system is already quite close to a normal distribution in its bulk when $\tau = 8 \tau_o$ -- which otherwise is asymmetric following the detailed form of the fluctuations in the transmission function encoded in the Voigt profile (see Ref.~\citenum{ochoa2019generalized})\footnote{More precisely, we notice that the absolute value of the pointwise difference between the normalized histogram $H$ and the Gaussian distribution $G$ with the same mean and variance, decreases as the sampling time increases. The ratio 
\[
q = \frac{\sum_i \big(H(I_i^*) - G(I_i^*)\big)^2}{\sum_i H(I_i^*)^2+G(I_i^*)^2},
\]
 where $I_i^*$ is the midpoint in the $i$th bin, can be used to quantify how different is the current distribution from the Gaussian limit. For $\tau = 8 \tau_o$, this ratio is less than 3 \%  within two standard deviations from the center for the model in Fig.\ \ref{fig:histogram}.}. In Fig.~\ref{fig:SNR_sampling}, we show the SNR as a function of the Fermi energy and for several sampling times. The enhancement in the SNR observed for larger sampling times, follows the trend anticipated in Eq.~\eqref{eq:SNR_1st_order}, i.e., it is proportional to  $\sqrt{\tau/\tau_o}$. As we will find in the next section, the tails approach to a limit value proportional to $1/\sigma_{\rm S}$, with a proportionality constant increasing with sampling time. The position at the maximum in the SNR, already identified in the Gaussian model Eqs.~\eqref{eq:MaxSNRGauss} and \eqref{eq:MaxSNRGausslim}, does not significantly change with the sampling time for this model. This result follows from the observation that the location of the maxima depends on the total broadening, which also includes the effect of the bias window, and the latter is not modified by rescaling. For the model and the sensing protocol investigated here, $\sigma_{\rm BW} \gg \sigma_{\rm S}$ such that total broadening  $\sigma \approx \sigma_{\rm BW}$.

%%%%%%%%%%%%%%%%%%%%%Fig1b%%%%%%%%%%%%%%%%%%%%%%%%
%%%%%%%%%%%%%%%%%%%%%%%%%%%%%%%%%%%%%%%%%%%%%%%%% 

\vspace{-0.5cm}
More generally, the sampling time will progressively washout the influence of the algebraic tails\footnote{This follows from the central limit theorem: For $S_n = \sum_i^n I_i$, $\left <  S_n \right> = n \bar I$. The characteristic function $\phi_{Z_n}(t)$, for the random variable $Z_n  = (S_n - n \bar I)/ \sqrt{n \sigma_I^2}$ equals to $\left< \exp [i t (I_1 - \bar I )/(\sqrt{n} \sigma_I)]\right>^n$. For large sampling times ($n \to \infty$), $\phi_{Z_n}(t) \to (1 - (1/2)t^2/n)^n \to e^{-t^2/2}$, which shows that $\phi_{Z_n}(t)$ converges to the characteristic function of a standard normal distribution process. By the L\'evy's continuity theorem $Z_n$ will be normally distributed, as well as $I_\tau$ with a mean value $\bar I$ and variance $(1/n)\sigma_I^2$.} in the current distribution.  Thus, the Gaussian model in Sec.~\ref{sec:gauss} is a good representation of the current distribution near its average value and for large sampling times relative to $\tau_o$. For a suspended graphene nanoribbon, $\tau_o$ is given by the relaxation time to a new independent configuration. In Ref.\ \citenum{gruss2018graphene}, such time is found to vary between 80 ps and 190 ps for a nanoribbon of 15~nm~$\times$~10~nm immersed in a water solution.  We will further verify that the Gaussian  model provides a good qualitative description of the proposed sensing protocol in the next section, but will also find that this approximation fails to accurately predict the form of the SNR.

\vspace{-0.4cm}
\section{ Approximate Voigt forms}\label{sec:lorentzian}
\vspace{-0.1cm}
In this section we examine optimal protocols for sensing energy shifts on a single level, taking into account the algebraic expression for the transmission function and its thermally broadened Voigt form\cite{gruss2018graphene,ochoa2019generalized}. Complementary to our previous work in Ref.~\citenum{ochoa2019generalized}, here we provide approximate analytical expressions for the current, noise, and electromechanical susceptibility.

The Taylor series of the current functional $I(\varepsilon_p)$ around the equilibrium energy $\bar \varepsilon_p$ can be used to approximate $\langle I \rangle$ as well as $\sigma_I$ (see Ref.~\citenum{ochoa2019generalized}).  Importantly, this approach leads to improved results as we include more terms in the expansion. In the case that $\varepsilon_p$ is normally distributed as in Eq.~\eqref{eq:g_e}, the moments of the distribution Eq.~\eqref{eq:g_e} are entirely determined by $\sigma_{\rm S}$ ($n \geq 1$):
\begin{align}
  \langle(\varepsilon_p - \bar \varepsilon_p)^{(2 n-1)}\rangle_g ={}& 0\\
  \langle(\varepsilon_p - \bar \varepsilon_p)^{2n}\rangle_g ={}& (2 n-1)!!\, \sigma_{\rm S}^{2n}.
\end{align}
 It follows that up to second order in $\sigma_{\rm S}$
\begin{align}
  \langle I \rangle  ={}& I (\bar \varepsilon_p) + \frac{\sigma_{\rm S}^2}{2}\,\partial_{\varepsilon_p}^2I(\bar \varepsilon_p) ,\label{eq:ILorentzian}\\
  \sigma_I^2  ={}& \sigma_{\rm S}^2 \left[ \left(\partial_{\varepsilon_p} I(\bar \varepsilon_p)\right)^2  + \frac{1}{2} \left( \partial_{\varepsilon_p}^2 I(\bar \varepsilon_p) \right)^2 \sigma_{\rm S}^2\right]\\
={}& \sigma_{\rm S}^2 \left[\left(\partial_{\varepsilon_p} I (\bar \varepsilon _p) \right)^2  + 2 (\langle I \rangle - I(\bar \varepsilon_p))^2\right]. \label{eq:variance_lorentzian}
\end{align}

We also observe that $\langle I \rangle \to I(\bar \varepsilon_p)$, and $\sigma_I^2 \to 0$ as $\sigma_{\rm S}^2 \to 0$. This shows that this variance captures the excess current noise due to mechanical fluctuations. We calculate $I(\varepsilon_p)$ from the exact form of the transmission function
\begin{align}
  T(\varepsilon - \varepsilon_p) ={}& \frac{w^2}{(\varepsilon - \varepsilon_p)^2 + w^2}, \label{eq:T_lorentzian}
\end{align}
for a single level coupled to two reservoirs with strength $w$. Utilizing the Gaussian approximation to the bias window, Eq.~\eqref{eq:Bias}, and writing $T$ in Eq.~\eqref{eq:T_lorentzian} as a partial fraction expansion (see Ref.~\citenum{ochoa2019generalized}), we obtain 
\begin{align}
  I(\varepsilon_p) ={}& I_{\rm V} \sigma_{\rm BW} {\rm Re}\left[\left(\sqrt{\frac{\pi}{2 \sigma_{\rm BW}^2}}-J(E, \sigma_{\rm BW}) \right)e^{\frac{E^2}{2 \sigma_{\rm BW}^2}} \right],\label{eq:Io_Lorentzian}
\end{align}
with
\begin{align}
  I_{\rm V} ={}& \frac{2 e}{h}\frac{w}{\sqrt{2 \pi}}\tanh\left(\frac{\beta \Delta \mu}{4} \right),
\end{align}
$E = w + i(\varepsilon_p - \mu)$ , $\bar E = w + i(\bar \varepsilon_p - \mu)$, and 
\begin{align}
  J(E,\sigma) = \frac{E}{\sigma^2}\int_0^1 d \alpha e^{-\frac{\alpha^2 E^2}{2 \sigma^2}} = \sqrt{\frac{\pi}{2}} {\rm erf}\left( \frac{E}{\sqrt{2}\sigma}\right).\label{eq:error}
\end{align}
The current in Eq.~\eqref{eq:Io_Lorentzian} takes the form of a Voigt profile in terms of the Fermi energy $\mu$, and consequently, should decay algebraically far from the peak maximum $\mu = \bar \varepsilon_p$. This is in contrast with the result for the model in Sec.~\ref{sec:gauss} in Eq.~\eqref{eq:Io_gauss}, in which case the decay is Gaussian. In terms of the bias window $\Delta \mu$, the currents in Eqs.~\eqref{eq:Io_gauss} and \eqref{eq:Io_Lorentzian} qualitatively agree, as the current amplitudes $I_{\rm G}$ and $I_{\rm V}$ have the same functional form. In particular, $I_{\rm G}$ and $I_{\rm V}$ coincide when the bias window dominates the fluctuations (i.e., $\sigma \approx \sigma_{\rm BW}$) and if $\sigma_T = w$.

To compute $\langle I \rangle$ from \eqref{eq:Io_Lorentzian}, utilizing Eq.~\eqref{eq:ILorentzian}, we notice that\footnote{In the verification of the last two expressions, the following identity was utilized
\begin{align*}
  \frac{d}{dx} J(x,\sigma) ={}& \frac{1}{\sigma^2} e^{-\frac{x^2}{2 \sigma^2}},  
\end{align*}
 }
\begin{align}
  \frac{\partial I(\varepsilon_p)}{\partial \varepsilon_p} ={}&  \frac{1}{\sigma_{\rm BW}^2} \Sigma_2(\varepsilon_p)\\
  \frac{\partial^2 I(\varepsilon_p)}{\partial \varepsilon_p^2} ={}& \frac{1}{\sigma_{\rm BW}^4} \Sigma_4(\varepsilon_p) ,
\end{align}
{\onecolumngrid \begin{center}
  \begin{figure}[t]
    \centering
    \includegraphics[scale=0.43]{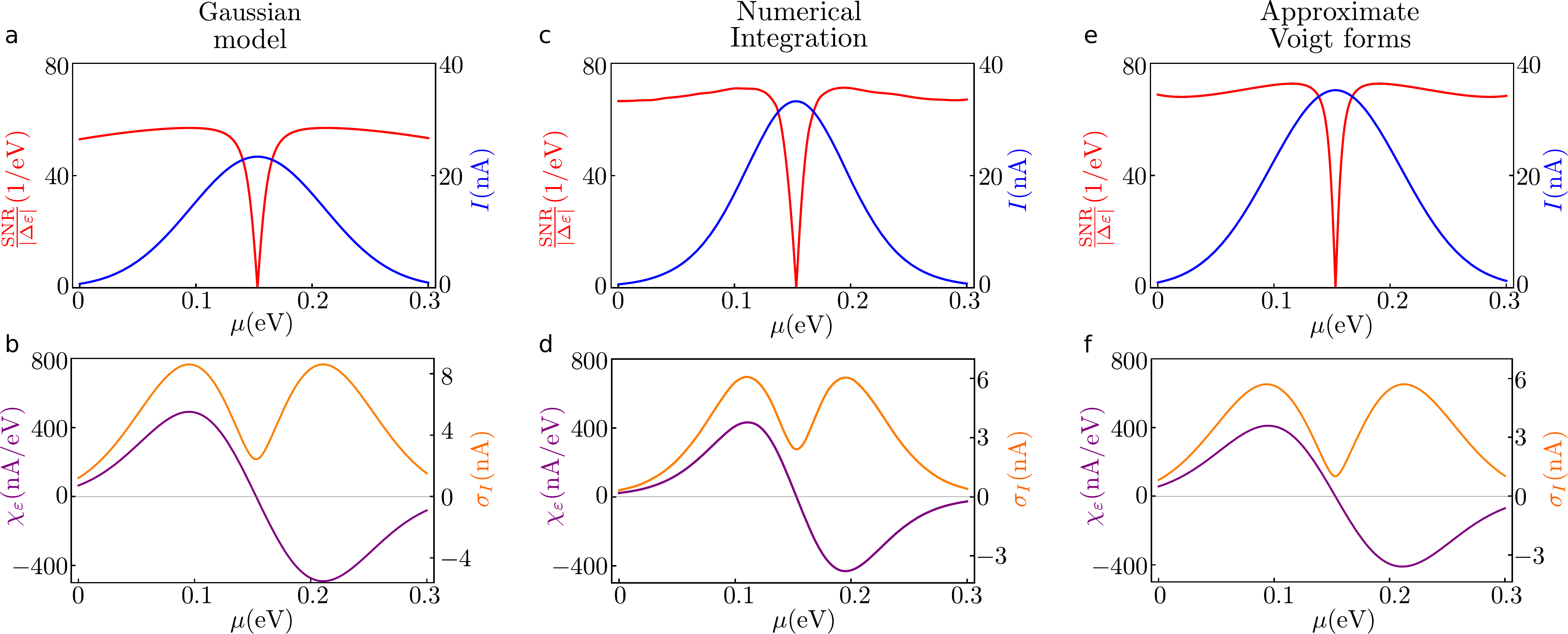} 
    \caption{(Color online) Sensing protocol for energy level shifts, analyzed with a Gaussian (a,b), numerically exact (c,d) and Voigt (e,f) estimates, for the system in Fig.~\ref{fig:histogram}. (a,c,e) Current (blue) and SNR (red) as a function of the Fermi level $\mu$ at a fixed symmetric bias of $\Delta \mu = 50$~meV. (b,d,f) Current variance (orange) and electromechanical susceptibility (purple) as a function of Fermi level $\mu$.  Parameters for this model are taken from Ref.~\citenum{ochoa2019generalized} and Fig.~\ref{fig:histogram}, and are representative of suspended graphene nanoribbons. $\bar \varepsilon_p =0.153$~eV, $w = 1.3$~meV, and 300~$K$. For the Gaussian model we use the parameters that fit the maximum and width at half maximum of the transmission function, i.e., $A=1$ and $\sigma_T = w /\sqrt{ 2 \ln 2} \approx 0.85 w$ (other parameters for the Gaussian model can be used depending on what properties one wants to reproduce, such as using $\sigma_T = w$ to match the current fluctuations).}
    \label{fig:sensingONE}
  \end{figure}
\end{center}
}\twocolumngrid

\noindent where we have introduced the coefficients $\Sigma_{2 n}$, which are proportional to the $n$th derivative of the current with respect to $\varepsilon_p$, with proportionality constant $1/\sigma^{2 n}_{\rm BW}$. Explicitly 
\begin{align}
  \Sigma_2(\varepsilon_p) ={}&  \; (\varepsilon_p - \mu) I(\varepsilon_p)-w K(\varepsilon_p) \label{eq:S2}\\
  \Sigma_4(\varepsilon_p) ={}&  \; w \, \sigma_{\rm BW}\, I_{\rm V} + 2 w (\varepsilon_p - \mu)K(\varepsilon_p) \notag\\
  &\hspace{1.5cm} - (\sigma_{\rm BW}^2+{\rm Re}[\bar E^2]) I(\varepsilon_p),\label{eq:S4}
\end{align}
with the auxiliary function $K(\varepsilon_p)$  defined as the imaginary counterpart\footnote{
  Explicitly
\begin{multline}
  K(\varepsilon_p) = I_{\rm V}  \sigma_{\rm BW} {\rm Im}\left[\left(\sqrt{\frac{\pi}{2 \sigma_{\rm BW}^2}}-J(E, \sigma_{\rm BW}) \right)e^{\frac{E^2}{2 \sigma_{\rm BW}^2}} \right].
\end{multline}
} of $I(\varepsilon_p)$ in Eq.~\eqref{eq:Io_Lorentzian}. Therefore, the stationary current under thermal fluctuations is
\begin{align}\label{eq:Itilde_Lorentzian}
  \langle I \rangle  ={}&  I(\bar \varepsilon_p)+ \frac{1}{2}\frac{\sigma_S^2}{\sigma_{\rm BW}^4} \Sigma_4(\bar \varepsilon_p).
\end{align}
Likewise, the contribution to the current variance due to environmental noise, Eq.~\eqref{eq:variance_lorentzian}, is 
\begin{align}
  \sigma_I^2 ={}& \frac{\sigma_{\rm S}^2}{\sigma_{\rm BW}^4} \Sigma_2^2(\varepsilon_p)+ \frac{1}{2}\frac{\sigma_{\rm S}^4}{\sigma_{\rm BW}^8} \Sigma_4^2(\varepsilon_p)  \label{eq:sigmaI_L}.
\end{align}
 We must emphasize that in the derivation of Eqs.~\eqref{eq:Itilde_Lorentzian} and \eqref{eq:sigmaI_L} we utilized only two approximations: the Gaussian form for the bias window in Eq.~\eqref{eq:Bias}, and the truncated Taylor series Eqs.~\eqref{eq:ILorentzian}-\eqref{eq:variance_lorentzian}.  These expansions for the thermally broadened current and variance up to second order in $\sigma_{\rm S}^2$ are accurate whenever thermal fluctuations are small compared with the bias window, i.e., $\sigma_{\rm S} < \sigma_{\rm BW}$, as is the case for the system investigated in Fig.~\ref{fig:histogram}.

 Next we provide an analytic expression for the susceptibility $\chi_\varepsilon$ that we find  by considering linear deviations in the current in Eqs.~\eqref{eq:Io_Lorentzian} and \eqref{eq:Itilde_Lorentzian}, due to a controlled level shift $\Delta \varepsilon$. The details of this derivation are presented in Appendix \ref{ap:susceptibility}, and the resulting approximate form is
\begin{multline}
  \chi_\varepsilon \approx \frac{ \Sigma_2(\bar \varepsilon_p)}{\sigma_{\rm BW}^2}\\
  + \frac{\sigma_{\rm S}^2}{ 2 \sigma_{BW}^4}\left [ \left(\frac{{\rm Re}[\bar E^2]}{\sigma_{\rm BW}^2} + 1\right)\Sigma_2 + 2 \frac{w (\bar \varepsilon_p - \mu)}{\sigma_{\rm BW}^2} \xi_2 \right ]_{\varepsilon_p = \bar \varepsilon_p},\label{eq:chi_lorentzian}
\end{multline}
where $\xi_2$ is defined by
\begin{equation}
  \label{eq:Sigma_tilde}
  \xi_2 = w I(\varepsilon_p) -(\varepsilon_p - \mu) K(\varepsilon_p).
\end{equation}

The expressions in Eqs.~\eqref{eq:sigmaI_L} and \eqref{eq:chi_lorentzian} for the current noise and susceptibility indicate that the SNR for the protocol here investigated reaches the limit value $|\Delta \varepsilon|/ \sigma_{\rm S}$, when the bias window dominates the fluctuations\footnote{This result follows calculating SNR with the lowest terms in the expansion for $\chi_\varepsilon$ and $\sigma_I$} (i.e., $\sigma_{\rm S} \ll \sigma_{\rm BW}$). We note that readout noise will have a more substantial effect when the Fermi level and the transmitting mode are well separated in energy, but this is far from the optimum setup that we find below. We also note that the variance $\sigma_I^2$ in Eq.~\eqref{eq:sigmaI_L} has an amplitude proportional to $I_{\rm V}^2$, while for the Gaussian model in Sec.~\ref{sec:gauss}, Eq.~\eqref{eq:sigmaIG} is proportional to $I_{\rm G}^2$. It follows that current fluctuations due to the noisy environment agree for both models when $\sigma_T = w$, and when the bias window is large.

We can now identify the signatures of inhomogeneous broadening in the current, current fluctuations and the SNR for the protocol investigated in Sec.~\ref{sec:gauss}. In Fig.~\ref{fig:sensingONE}, we compare these magnitudes for the model in Fig.~\ref{fig:histogram}, as obtained by numerical integration and with the analytical predictions for the fully Gaussian and the approximate Voigt forms for the transmission function. We observe that for this system $\sigma_{\rm S} < \sigma_{\rm BW}$. The approximate Voigt forms obtained in this section are valid in this regime.

\begin{center}
  \begin{figure}[t]
    \centering
    \includegraphics[scale=0.66]{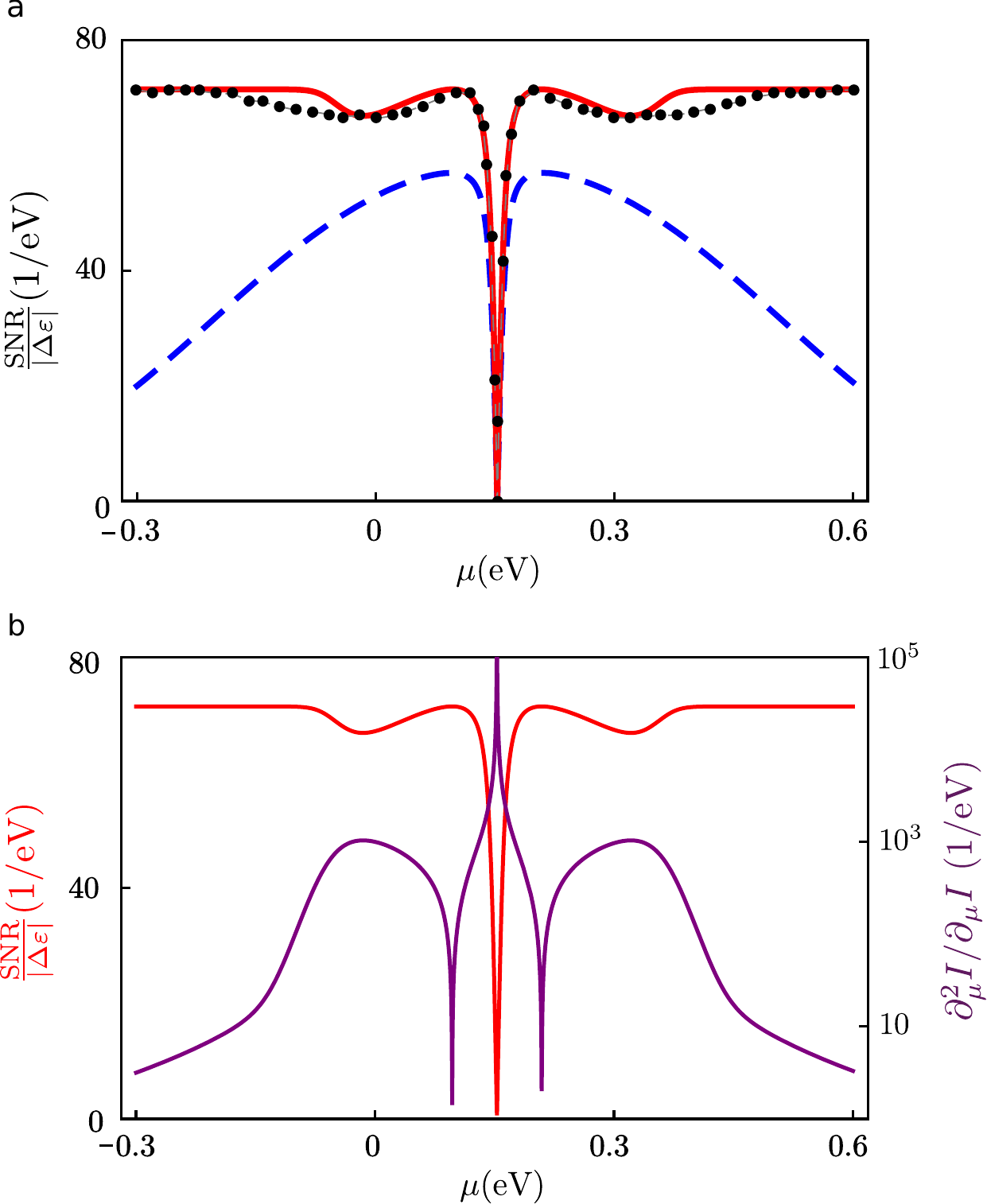} 
    \caption{(Color online) Signal-to-noise ratio SNR far from the peak.  (a) SNR for a the fully Gaussian (blue, dashed) and the Voigt (red, solid), and numerical integration (black, dots) for the model system investigated in Fig.~\ref{fig:sensingONE}. Notice that the SNR obtained from the Gaussian approximation to the transmission function decays to zero far from the main feature. In the case of the approximate Voigt forms the SNR achieves a constant nonzero value proportional to $1/\sigma_{\rm S}$, in agreement with the numerical result, albeit with a more rapid rise to the large $\mu$ SNR. This is due to the Gaussian approximation to the bias window in Eq.~\eqref{eq:Bias}. (b) SNR from the approximate Voigt forms and the ratio between the second and first derivatives of the current, showing that the maximum in the SNR occurs near $\mu = \varepsilon_p \pm \sigma_{\rm BW}$, and the minima in the tails correspond to local maxima in the derivative ratio near the point where the Voigt profile decays only algebraically.}
    \label{fig:SNRboth}
  \end{figure}
\end{center}

We also observe that most qualitative properties in the current, the linear response and the fluctuations near the main peak are captured already by the Gaussian approximation to the transmission function (Figs.~\ref{fig:sensingONE}b,d,f) discussed in Sec.~\ref{sec:gauss}. Deviations are due to the impossibility of fitting a Gaussian to a Lorentzian distribution with full accuracy, and because they have different decays far from its center $\bar \varepsilon_p$. This difference is manifested in the qualitative behavior  predicted for the SNR: the correct decay of the SNR from the main peak is only captured by the Voigt profile. Indeed, in the Gaussian picture $\chi_\varepsilon$ and $\sigma_I$ decay far from the point of current maximum as $e^{-\mu^2/(2 \sigma^2)}$ and $e^{-\mu^2/(2 (\sigma^2+\sigma_{\rm S}^2))}$, respectively. As a consequence, the SNR also decays as $e^{-\mu^2 \sigma_{\rm S}^2 /(\sigma^2 (\sigma^2+\sigma_{\rm S}^2))}$. On the contrary, the Voigt forms derived above decay algebraically and the SNR approaches asymptotically to a constant value proportional to $1/\sigma_{\rm S}$. This difference in the tails is shown in Fig.~\ref{fig:SNRboth}a, where both analytic forms for the SNR are contrasted with the exact result. The approximate Voigt forms decay faster than the numerically exact tails to the asymptotic value. This difference is due to the Gaussian approximation to the bias window: when the overlap between the bias window and the transmission function is small, this approximation underestimates the current and its response as we modulate $\mu$.  Figure~\ref{fig:SNRboth}a shows two maxima near the main depth for the SNR calculated from the approximate Voigt expressions.   We can understand the main characteristics in the SNR from the approximate Voigt form by writing
\begin{align}
  {\rm SNR} \approx{}& {\frac{1}{\sigma_{\rm S}}}\left(1 + \frac{1}{2}\frac{\sigma_{\rm S}^2}{\sigma_{\rm BW}^4} \frac{\Sigma_4^2}{\Sigma_2^2}\right)^{-1/2}.
\end{align}
First, we notice that a maximum in the SNR occurs when the ratio $|\Sigma_4/\Sigma_2|$ is minimal, and that this observation leads to the estimate $\mu_{\rm max} = \varepsilon_p \pm \sigma_{\rm BW}$\footnote{Notice that $|\Sigma_4/\Sigma_2| \sim |\partial_\mu^2 I/ \partial_\mu^2 I|$ and maximum values in $\Sigma_2$ correspond to zeroes in $\Sigma_4$. Next, notice that Eqs.~\eqref{eq:S2} and \eqref{eq:S4} are proportional to the Voigt forms $I$ and $K$. From the Gaussian component of this form, we obtain that $\Sigma_2$ is maximal at this values}. The SNR ratio then decays as the Gaussian component of the Voigt form and reaches a minimum value near the turnover point, where the algebraic decay determined by the error distribution dominates the decay. This is illustrated in Fig.~\ref{fig:SNRboth}b.

 %%%%%%%%%%%%%%%%%%%%%Fig3%%%%%%%%%%%%%%%%%%%%%%%%

%%%%%%%%%%%%%%%%%%%%%%%%%%%%%%%%%%%%%%%%%%%%%%%%%
\vspace{-1.0cm}
In summary, while qualitatively for the parameters in Fig.~\ref{fig:sensingONE}, either the Gaussian model or the approximate Voigt forms are reasonable, but the latter captures the SNR and current better, however, in other parameter regimes, in particular when the Fermi level is far from the transmitting mode, only the Voigt form captures the behavior of  the SNR.

\vspace{-0.6cm}
\section{conclusions}\label{sec:conclusion}
\vspace{-0.4cm}

We studied the electric current and fluctuations under inhomogeneous environmental conditions, providing analytical expressions for these quantities in two limiting cases. When the electronic transmission function is approximated by a Gaussian form, these magnitudes are Gaussian as well, with variance determined by the independent contribution of the coupling to the metal, the bias, and inhomogeneous conditions. On the contrary, starting from the exact rational form of the transmission function, the current takes a Voigt form. The Voigt lineshape is also imprinted in the behavior of the fluctuations. We also derived expressions for the electrical susceptibility and SNR in both cases and analyzed a protocol for optimal sensing. These results indicate that the algebraic decay in the Voigt forms, due to the inhomogeneous conditions, generally must be incorporated in the description and design of optimal sensing protocols, although approximations such as the Gaussian model can capture the proper behavior in particular parameter regions.

\vspace{-0.8cm}
\section{Supplementary Material}
\vspace{-0.5cm}
See supplementary material for an extended analysis of the Gaussian approximation to the bias window, current histograms as well as additional details on the derivation of the approximate Voigt forms.

\vspace{-0.8cm}
\begin{acknowledgments}
  \vspace{-0.5cm}
 M. A. O. acknowledges support under the Cooperative Research Agreement between the University of Maryland and the National Institute of Standards and Technology Physical Measurement Laboratory, Award 70NANB14H209, through the University of Maryland. 
\end{acknowledgments}

\appendix
\vspace{-0.8cm}
\section{Derivation of  SNR in Eq.~\eqref{eq:SNR_1st_order}}\label{ap:x_variance}
\vspace{-0.5cm}
For a Brownian particle $Y$ in a quadratic potential with mass $m$, $\bar Y = 0$, spring constant $\kappa$, and frequency $\omega_o$, the correlation function $C_Y(t)$ is given by\cite{zwanzig2001nonequilibrium}
\begin{equation}\label{eq:CXt}
C_Y(t) = \frac{1}{\kappa \beta} e^{-\frac{\eta}{2 m} t}\left( \cos \omega_1 t+ \frac{\eta}{2 m \omega_1} \sin \omega_1 t \right),
\end{equation}
with $\omega_1^2 = \omega_o^2 - (\eta/2m)^2$. Defining $y = (1/\tau) \int_0^{\tau} d t Y(t)$ and utilizing Eq.~\eqref{eq:CXt}  one can evaluate $ \left <  \delta y^2 \right> $ as follows. First notice that $\langle y \rangle = (1/\tau) \int_0^{\tau} d t \langle Y(t) \rangle = 0$, such that $\langle \delta y^2 \rangle =\langle y^2 \rangle $ and
\begin{align}
  \langle y^2 \rangle ={}& \frac{1}{\tau^2} \int_0 ^\tau dt \int_o^\tau dt' \langle Y(t) Y(t')\rangle, 
\end{align}
which can be written in terms of the variables $s = t - t'$ and $\alpha = (t+t')/2$ such that $\langle Y(t) Y(t')\rangle = C_Y(s)$. After integration with respect to the new variables we obtain
\begin{align}
  \langle y^2 \rangle ={}&\frac{\eta}{\tau \kappa^2 \beta} \notag\\
&+ \frac{e^{-\frac{\eta}{2 m}\tau}}{\tau \omega_1 \kappa \beta}\left[\sin (\omega_1 \tau)+\frac{\eta}{m}{\rm Re}\left(\frac{e^{i \omega_1 \tau}}{\omega+ i \eta/2 m} \right)  \right],
\end{align}
and we can disregard the second term in the case $\tau > \eta^{-1} m$. 

\vspace{-0.8cm}
\section{Sampling time, variance and skewness}\label{ap:skewness}
\vspace{-0.5cm}
Here we show that the third moment, decays as the inverse of the square of the sampling time.
First, for the variance
\begin{align}
  \left \langle(I_\tau - \bar I)^2 \right \rangle ={}& \frac{1}{n^2} \left \langle \left(\sum_i^n I_i - \bar I\right)^2\right \rangle\\
={}& \frac{1}{n^2}\left \langle \sum_i^n (I_i - \bar I)^2\right \rangle \notag\\
& + \frac{2}{n^2} \left \langle \sum_i \sum_{j>i} (I_i - \bar I)(I_j - \bar I)\right \rangle\\
={}& \frac{1}{n^2} \sum_i^n \sigma^2 = \frac{1}{n} \sigma^2.
\end{align}
For the skewness, we notice that
\begin{align}
  \left(\sum_i^n I_i - \bar I \right)^3 ={}& \sum_i^n (I_i - \bar I)^3 + 2 \sum_i^n (I_i- \bar I)^2 \sum_{j > i} (I_j - \bar I)\notag\\
& + 6 \sum_i^n  \sum_{j>i} \sum_{k>j} (I_i - \bar I ) (I_j - \bar I ) (I_k - \bar I ) 
\end{align}
The result in Eq.~\eqref{eq:skewN} follows from this result, and the fact the expectation $\langle \rangle$ is a linear function.

\vspace{-0.5cm}
\section{Electromechanical susceptibility $\chi_\varepsilon$}\label{ap:susceptibility}
\vspace{-0.5cm}
In this section we derive the electrochemical susceptibility $\chi_\varepsilon$. We begin by writing the instantaneous current in Eq.~\eqref{eq:Io_Lorentzian} for a shifted energy level $\varepsilon_p + \Delta \varepsilon$ 
  \begin{align}
    I(\varepsilon_p +{}& \Delta \varepsilon) = \frac{w \sigma_{\rm BW}}{\sqrt{2 \pi}}\tanh\left(\frac{\beta \Delta \mu}{4} \right) \times \notag\\
    &{\rm Re}\left[\left(\sqrt{\frac{\pi}{2 \sigma_{\rm BW}^2}}-J(E+i\Delta \varepsilon, \sigma_{\rm BW}) \right)e^{\frac{(E+i\Delta \varepsilon)^2}{2 \sigma_{\rm BW}^2}} \right],\label{eq:Idelta_full}
  \end{align}
 perform expansions in terms $\Delta \varepsilon$, and recover the linear terms in the level shift. For this, we utilize the approximations
\begin{align}
  (E+ i \Delta \varepsilon)^2 \approx{}& \; E^2 + 2 i E \Delta \varepsilon\\
e^{\frac{(E+i\Delta \varepsilon)^2}{2 \sigma_{\rm BW}^2}} \approx{}&\; e^{\frac{E^2}{2 \sigma_{\rm BW}^2}}\left(1+ i\frac{E \Delta \varepsilon}{\sigma_{\rm BW}^2} \right) \\
J(E+i\Delta \varepsilon, \sigma_{\rm BW}) \approx{}&\; J(E, \sigma_{\rm BW})+ i \Delta \varepsilon \frac{\partial}{\partial E}J(E, \sigma_{\rm BW}),
\end{align}
which hold for small shifts. As a result, we linear form of Eq.~\eqref{eq:Idelta_full} is 
\begin{align}
  I^{(1)}(\varepsilon_p +\Delta \varepsilon) ={}& I(\varepsilon_p) + \Delta \varepsilon \frac{\Sigma_2 }{\sigma_{\rm BW}^2}.\label{eq:DeltaI1}
\end{align}
In a similar fashion we find
\begin{align}
    K^{(1)}(\varepsilon_p +\Delta \varepsilon) ={}& K(\varepsilon_p) +\Delta \varepsilon \frac{\xi_2(\bar \varepsilon)}{\sigma_{\rm BW}^2}. \label{eq:DeltaK1}
\end{align}
with $\xi_2$ given by Eq.~\eqref{eq:Sigma_tilde}. Substituting Eqs.~\eqref{eq:DeltaI1} and \eqref{eq:DeltaK1} in Eq.~\eqref{eq:Itilde_Lorentzian}, and collecting only terms that are linear in $\Delta \varepsilon$ we obtain the expression in Eq.~\eqref{eq:chi_lorentzian}.

%An alternative form for $\chi_\varepsilon$ in terms of $\langle I \rangle$ is
% \begin{align}
%  \chi_\varepsilon \approx & (\bar \varepsilon_p -\mu) \times \notag\\
%  &\left[ \frac{\sigma_{\rm S}^2}{\sigma_{\rm BW}^2} I(\bar \varepsilon_p)+\frac{1}{\sigma_{\rm BW}^2}\left(\langle I \rangle  - \frac{w \sigma_{\rm BW}}{\sqrt{2 \pi}} \tanh \left(\frac{\beta \Delta \mu}{4} \right) \right) \right] \notag\\
%&\hspace{1cm} -\frac{w}{\sigma_{\rm BW}^2} \left(1 - \frac{1}{2}\frac{\sigma_{\rm S}^2}{\sigma_{\rm BW}^2}+ \sigma_{\rm S}^2\frac{{\rm Re}(\bar E ^2)}{2 \sigma_{\rm BW}^2} \right) K(\bar \varepsilon_p). %+ o(\Delta \varepsilon^2)
%\end{align}

%\section{Approximate For for the Voigt profile, current and Noise}\label{ap:VoigtCurrent}
%\begin{equation}
%  \int_0^1 e^{-\frac{\alpha^2 x^2}{2 \sigma^2}} = \frac{2 \sigma^2}{x^2}\left(e^{-\frac{ x^2}{2 \sigma^2}}-1\right) + \sum_k
%\end{equation}

\nocite{*}
\bibliography{RefOptimal2.bib}% Produces the bibliography via BibTeX.

\end{document}

\preprint{AIP/123-QED}

\title{Supplementary Material -- Optimal {\itshape in situ} electromechanical sensing of molecular species}% Force line breaks with \\
%\thanks{Footnote to title of article.}

\author{Maicol A. Ochoa}
\affiliation{Biophysics Group, Microsystems and Nanotechnology Division, Physical Measurement Laboratory, National Institute of Standards and Technology, Gaithersburg, MD 20899}
\affiliation{Maryland Nanocenter, University of Maryland, College
Park, MD 20742}

\author{Michael Zwolak}
\email{mpz@nist.gov}
\affiliation{Biophysics Group, Microsystems and Nanotechnology Division, Physical Measurement Laboratory, National Institute of Standards and Technology, Gaithersburg, MD 20899}
 %\homepage{http://www.Second.institution.edu/~Charlie.Author.}
%\affiliation{%

%Second institution and/or address%\\This line break forced% with \\
%}

\date{\today}% It is always \today, today,

\maketitle

\section{Transmission Function, Voigt profile, and Bias Window }

\begin{figure}[tb]
  \centering
  \includegraphics[scale=0.65]{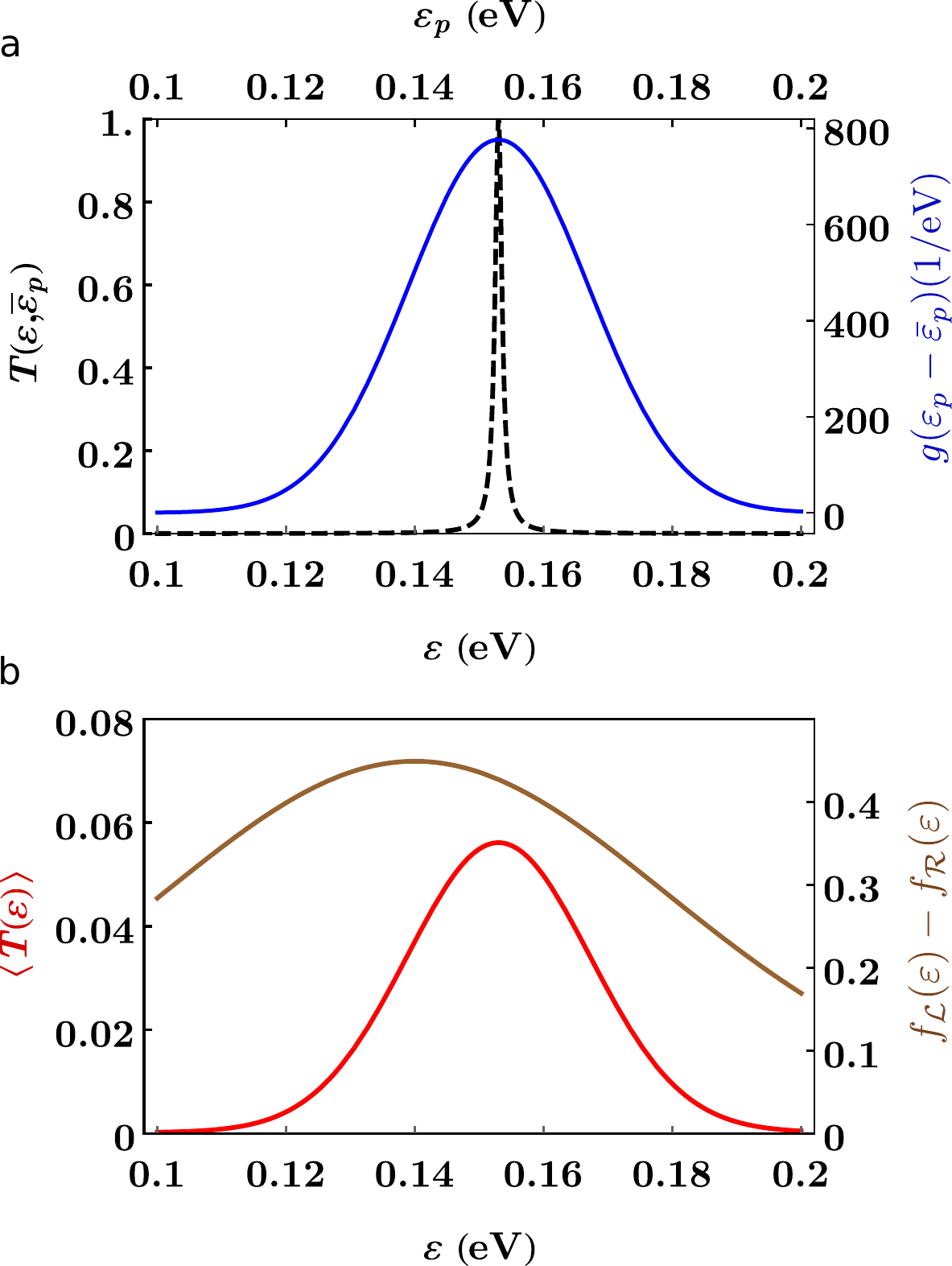}
  \caption{Transmission electronic function and Bias window for a single level coupled to two fermionic reservoirs.(a) The Lorentzian transmission function $T$ (black, dashed) and the normal distribution of eigenvalues $\varepsilon_p$ (red, solid). (b) The Voigt profile $\langle T \rangle $ (Red, solid) resulting from the convolution of $T$ and $g$, and the bias window (brown, solid) for a symmetric bias at 300~K. Parameters for this model are $\bar \varepsilon_p = 0.153$~eV, $w = 1.3$~meV, $\sigma_{\rm S} = 14$~meV, $\mu = 0.14$~eV and $\Delta \mu = 50$~meV. \label{fig:Voigt}}
\end{figure}

 \begin{figure}[hbt]
  \centering
  \includegraphics[scale=0.385]{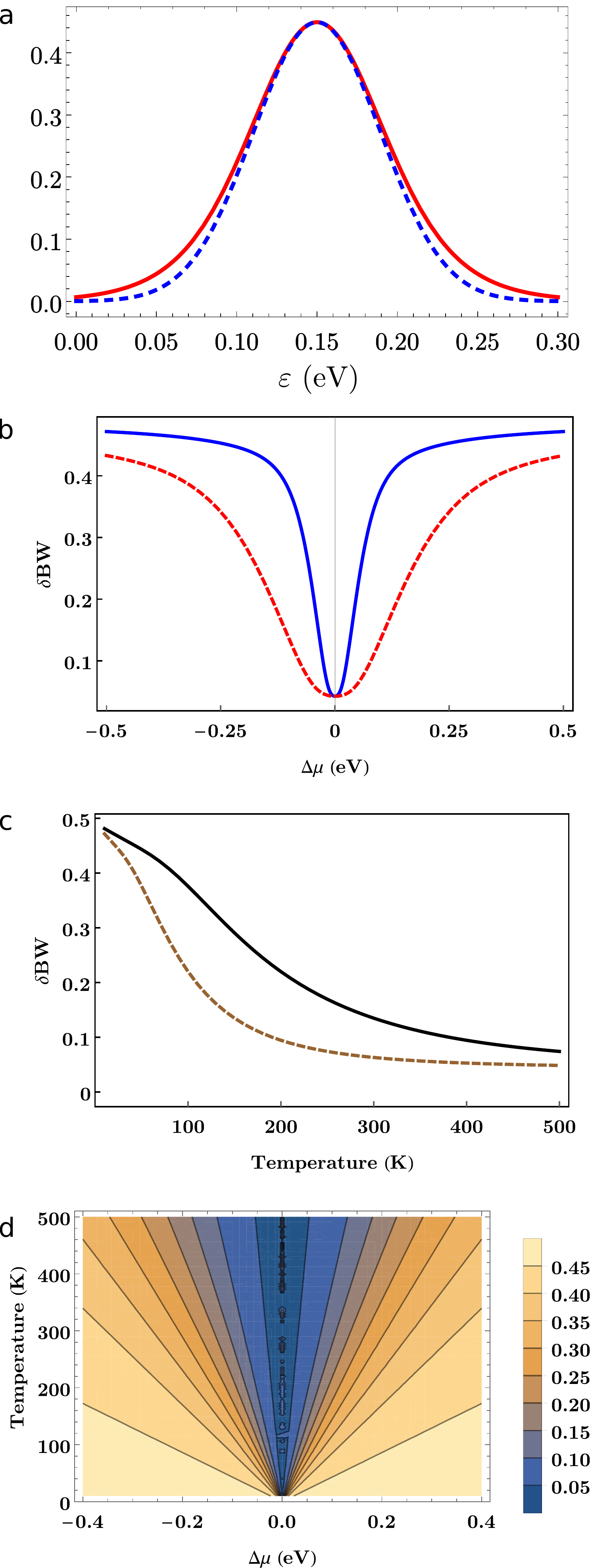}
  \caption{Bias Window $f_{\mathcal{L}}(\varepsilon)-f_{\mathcal{R}}(\varepsilon)$ and the Gaussian approximation Eq.~(6). (a) A numerical comparison of the exact (red, solid) and the approximate (blued, dashed) forms at 300~K, $\Delta \mu =50$~meV and $\mu = 0.153$~eV. (b) Normalized difference $\delta \rm BW $ as a function of the difference in chemical potential $\Delta \mu$ at 100~K (blue, solid) and 300~K (red, dashed). (c) $\delta \rm BW$ as a function of the contact temperature for $\Delta \mu = 0.1$~eV (Black, solid) and $\Delta \mu = 0.05$~eV. d) Contour plot for $\delta \rm BW$. These figures suggests that the Gaussian approximation to the Bias Window in Eq.~(6) on the main text provides a good estimate of the bias window near the Fermi energy $\mu$, for small $\Delta \mu$ and large Temperature. \label{fig:BW}}
\end{figure}

As a result of the environmental fluctuations, the electronic transmission properties of suspended graphene nanoribbons experience inhomogeneous broadening, in addition to the smearing imprinted by the metallic contacts. In case the fluctuations are slow relative to the time scale for electron transport, a separation of nuclear and electronic degrees of freedom is possible in the form of the Born-Oppenheimer approximation (see Ref.~18). The effective transmission function $\langle T \rangle$ is therefore the thermal average of the electronic transmission function $T$ over the multiple configurations accessible to the graphene matrix.  If the fluctuations follow a Gaussian distribution such as in Eq.~(2), $\langle T \rangle$ takes the form of a Voigt and Generalized Voigt profile, whenever $T$ is a Lorentzian or rational form, respectively. In Fig.~\ref{fig:Voigt}, we compare the static $T(\varepsilon, \bar \varepsilon_p)$ and thermally broadened (Voigt) transmission functions $\langle T (\varepsilon) \rangle$ for the particular case of a single level $\varepsilon_p$ coupled to two fermionic reservoirs. We also show the profile for the bias window, $f_{\mathcal{L}}(\varepsilon)- f_{\mathcal{R}}(\varepsilon)$, and the normal distribution for the energy level $\varepsilon_p$ at 300~K, with a standard deviation of 14~meV and centered at $\bar \varepsilon_p = 0.153$~eV. 

\section{ Gaussian approximation to the Bias Window. }

\begin{figure}[htb]
  \centering
  \includegraphics[scale=0.5]{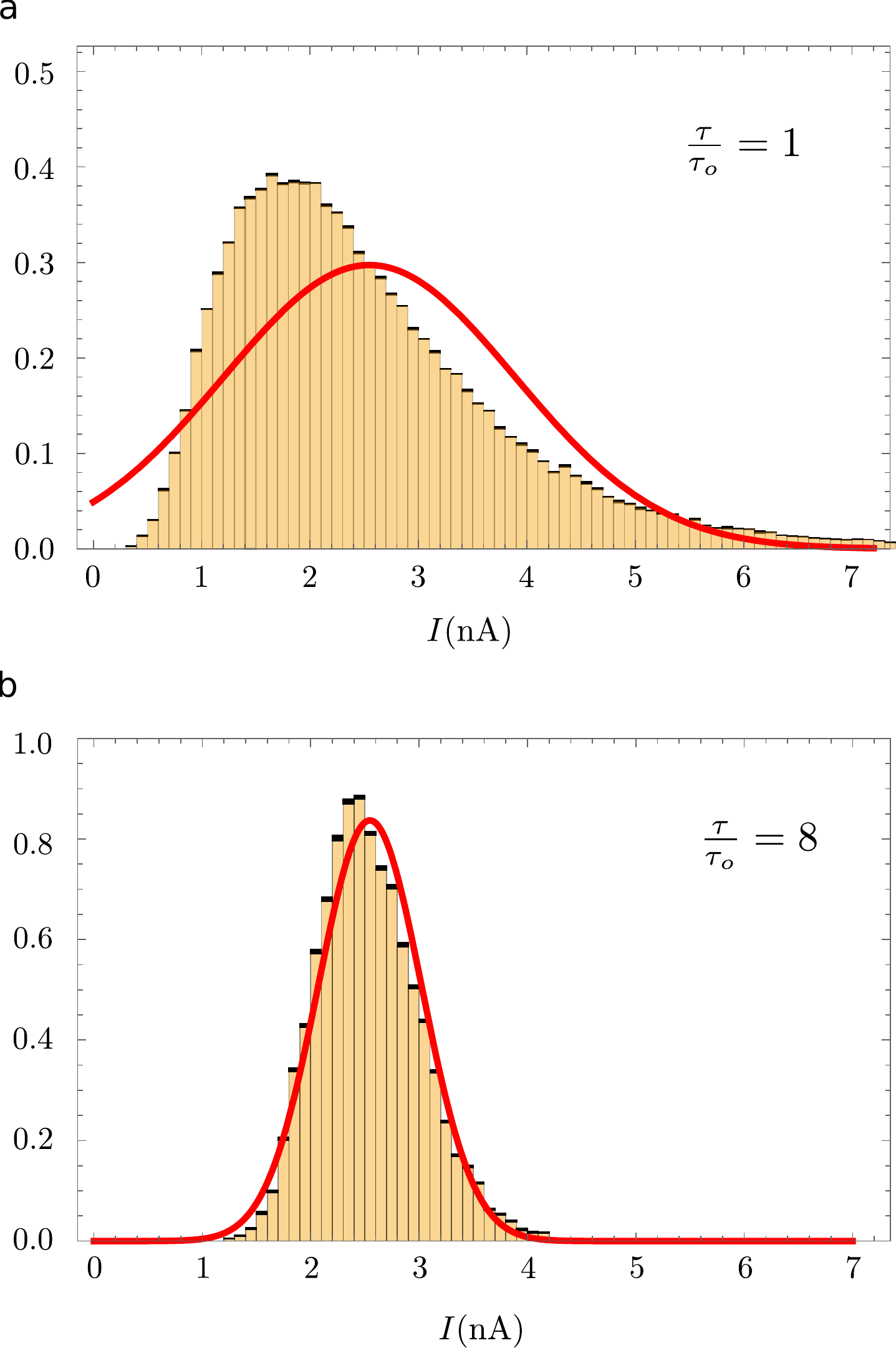}
  \caption{Normalized current histograms for sampling times (a) $\tau_o$ and  (b) $8 \tau_o$ for a sample size of 80000 points. The number of bins and with is determined by the Scott's normal reference rule and corresponds to 0.1~nA. The solid line is the normal density function with variance and mean value of the corresponding histogram. The standard deviation for each bin is of the size of the thickness of bin borderline. \label{fig:hist} }
\end{figure}

Electron transport through suspended graphene nanoribbons is possible due to the coupling between the graphene electronic modes and those of the metals, and is driven by the drop in potential energy experienced by an electron transferred between the contacts. This driving force is determined by bias window: the difference in equilibrium electron distribution between the contacts $f_{\mathcal{L}}(\varepsilon)- f_{\mathcal{R}}(\varepsilon)$, where $f_{\mathcal{L}/\mathcal{R}}$ denotes the Fermi function for the left/right metal. A bias window can be induced by, for example, the difference in chemical potential $\Delta \mu = \mu_{\mathcal{L}}- \mu_{\mathcal{R}}$. Figure \ref{fig:Voigt}b shows a particular realization of the bias window near room temperature with $\Delta \mu = 50$~meV.  The Fermi function $f$ varies smoothly near the Fermi energy $\mu$, and the difference in the electron distribution in the contacts has a bell shape. We introduce a Gaussian approximation $G_{\rm BW}$ to the bias window and assess the accuracy of this approximation. A Gaussian function is defined to match the maximum and the width at half maximum of $f_{\mathcal{L}}- f_{\mathcal{R}}$. The resulting expression is Eq.~(6). In Fig.~\ref{fig:BW} we compare the exact and approximate forms, and show how the Gaussian approximation to the bias window performs at different temperatures and gradient in chemical potential.

In order to quantify the accuracy of the approximation, we introduce the factor $\delta \rm BW$
\begin{equation}
  \label{eq:dBW}
  \delta {\rm BW} = \frac{|| f_{\mathcal{L}} - f_{\mathcal{R}} - G_{\rm BW} ||}{|| f_{\mathcal{L}}-f_{\mathcal{R}}||+||G_{\rm BW} ||} \tag{S1}
\end{equation}
where $G_{\rm BW}(\varepsilon)$ is the Gaussian function in Eq.~(6), and $|| \cdot ||$ is the $L^2(\mathbb{R})$ norm defined by 
\begin{align}
  || \phi || = \left( \int_{-\infty}^{\infty} d\varepsilon\, \phi(\varepsilon)^2 \right)^{1/2} \tag{S2}
\end{align}
for an $L^2$ function $\phi$. The factor $\delta \rm BW$ defined in Eq.~\eqref{eq:dBW} varies from zero to one for rapidly decaying functions, and vanishes in the ideal case that our approximation $G_{\rm BW}$ equals at every point the actual form of the Bias window. Thus, this factor provides a quantitative measure of the accuracy of the Gaussian approximation to the bias window. In Fig.~\ref{fig:BW}b-d, we investigate $\delta \rm BW$ as a function of $\Delta \mu$ and temperature. We observe that the approximation $G_{\rm BW}$ introduced in Eq.~(6) is accurate at high temperatures and small gradients in chemical potential. In particular, for the protocol investigated in this manuscript $\delta \rm BW \approx 0.063$. It is possible to introduce other measures to account for the accuracy of this approximation in terms of other norms for the space of density functions.

\section{Histograms and the Gaussian model}

In this section we illustrate how coarse graining in time brings the current distribution into a Gaussian form. This observation was already discussed in Fig.~1 in the main text, and we provide a more quantitative account of this here. Figure \ref{fig:hist} shows the normalized current histogram for a sample size of 80000 points, with $\mu =0.08$~eV, for two different sampling times $\tau$, defined relative to the intrinsic time $\tau_o$. The histogram normalization guarantees that the sum over the histogram frequency multiplied by the bin size equals unity. This permits a direct comparison with the normal distribution of identical mean value and variance.  The number of bins and width are determined by the Scott's normal reference rule that minimizes the integrated mean squared error. We observe in Fig.~\ref{fig:hist} that the normalized histogram follows closely the corresponding normal distribution when $\tau = 8 \tau_o$, and fails to reproduce the asymmetric form of the current histogram for a sampling time $\tau = \tau_o$. In fact, utilizing the approach described in the Ref. [36], we obtain a fit factor $q$ of $0.7$~\% in the case of $\tau = 8 \tau_o$, while for $\tau = \tau_o$ we obtain a $q = 6.3$~\%. 

\begin{figure}[tbh]
  \centering
  \includegraphics[scale=0.55]{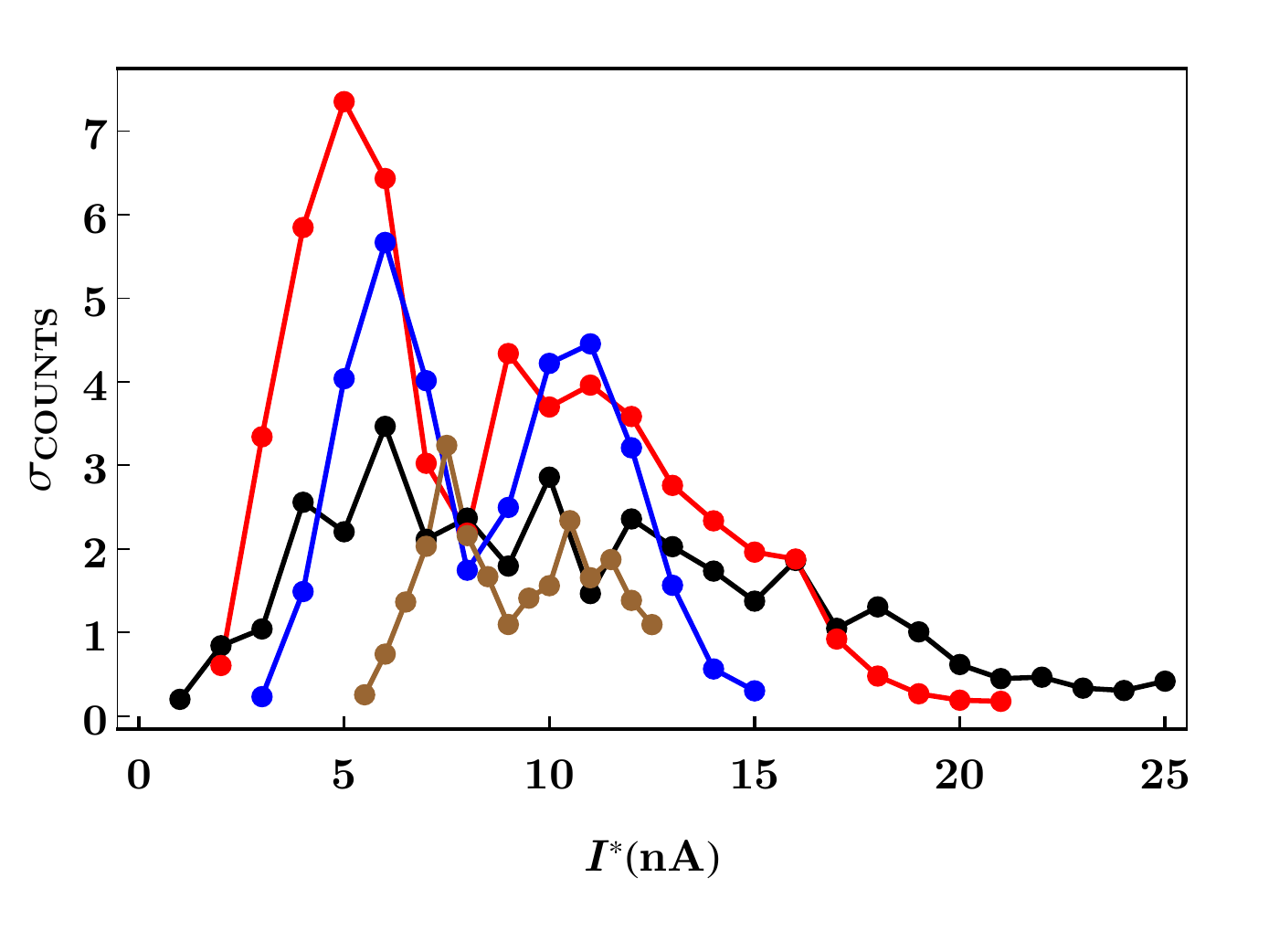}
  \caption{Standard deviation in the counts per bin for the four current histograms in Fig.~1. Each histogram in Fig.~1 differs in the sampling time $\tau$, and their corresponding counting error are (black) $\tau = \tau_o$, (red) $ \tau = 2 \tau_o$, (blue) $\tau = 4 \tau_o$, and  (brown) $\tau = 8 \tau_o$.   \label{fig:hist_errors}}
\end{figure}

Current histograms constructed from a sample of size $n$ and consisting of $k$ bins, carry an uncertainty in the counts per bin that we quantify with the standard error $\sigma_i$, $i = 1,\dots,k$. If $n$ is large enough, the standard error in the counts per bin can be calculated by the following method. First, we partition the original sample into $10$ subsamples of equal size ( $n/10$ ). For each subsample we generate a histogram with $k$ bins, and evaluate for each bin the standard deviation in the counts from these 10 histograms. Then, we rescale the standard deviation and obtain its corresponding value for the original histogram by dividing  each $\sigma_i$ by $\sqrt{10}$. This renormalization assumes that the variance in the counts per bin is inversely proportional to the sample size, which requires that $n$ is large. In this form, we obtained the standard deviation for the histograms in Fig.~1, which are shown independently in Fig.~\ref{fig:hist_errors}.

\section{Approximate Voigt form in Eq. (33)}

The approximate Voigt forms in Sec.~IV result from the analytic integration of the static current integral utilizing the Gaussian approximation to the bias window in Eq.~(6). The general procedure is essentially described in Ref.~[18], and we repeat the argument for the case of the instant current in Eq. (33).
First we write  the transmission function $T$ as
\begin{align}
T(\varepsilon, \varepsilon_p) =\frac{w}{2}\left( \frac{1}{E} +\frac{1}{E^*}\right) \tag{S3}
\end{align}
 where $E = i(\varepsilon - \bar \varepsilon_p) +w $. Then, we write each fraction as an integral of an exponential 
\begin{align}\label{eq:apEinv}
  \frac{1}{E} =& \int_0^\infty da\; e^{-a E}. \tag {S4}
\end{align}
In this form the Landauer-B\: uttiker formula for the instant current reads
\begin{align}
  I(\varepsilon_p) =&  \frac{w}{2} \int_0^\infty da \int \frac{d\varepsilon}{2 \pi} \left( e^{-a E} + e^{-a E^*}\right) (f_{\mathcal{L}}(\varepsilon)-f_{\mathcal{R}}(\varepsilon)), \tag{S5}
\end{align}
which can be integrated with the aid of the Gaussian approximation to the Bias window in Eq.~(6).

\begin{figure}[hbt]
  \centering
  \includegraphics[scale=0.55]{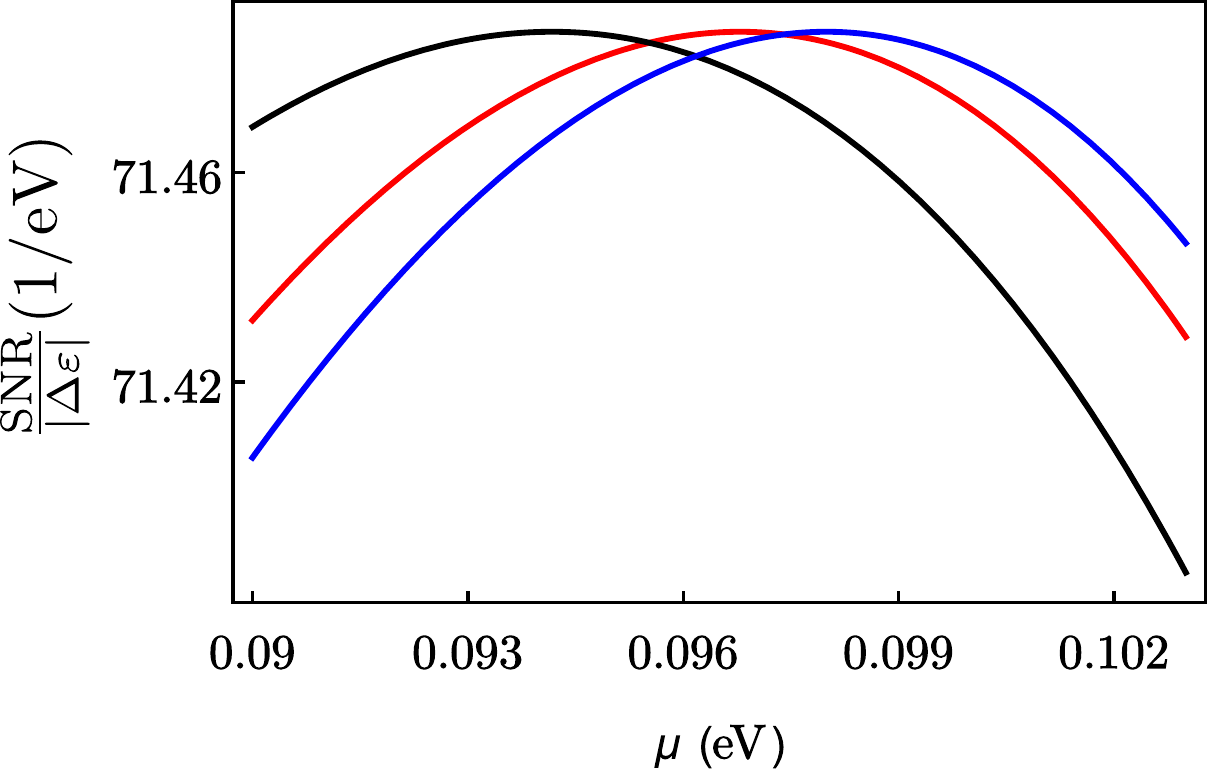}
  \caption{Signal-to-noise ratio, near the local maximum $\mu_{\rm max}$ for three different gradients in chemical potential $\Delta \mu$. Parameters for this model are the same as in Fig.~4, with (black) $\Delta \mu =$ 100~meV, (red) $\Delta \mu =$ 50~meV, and (blue) $\Delta \mu =$ 25 meV.  }
  \label{fig:Max}
\end{figure}

\section{Local maxima in the SNR, Fig.~4. }

 The approximate form for the SNR in Eq.~(44) and Fig.~4, reveals two local maxima at $\mu_{\rm max} = \bar \varepsilon_p \pm \sigma_{\rm BW}$, with $\sigma_{\rm BW}$ functionally dependent on the gradient in chemical potential $\Delta \mu$ (see Eq.~(6)). When the product $\Delta \mu \beta$ is bigger than 16, $\sigma_{\rm BW}$ is well approximated by $\sigma_{\rm BW}  \approx \Delta \mu / (4 \sqrt{2 \ln 2} )$. In such regime, $\sigma_{\rm BW}$ and $\mu_{\rm max}$ are independent of the temperature. However, at room temperature (300~K) such conditions will be satisfied if $\Delta \mu \geq 400$~meV, and the Gaussian approximation to the bias window Eq.~(6) breaks as indicated by the Fig.~\ref{fig:BW}d. The maximum in SNR as a function of the Fermi energy can be modulated by the varying $\Delta \mu$, but the change, as illustrated in Fig.~\ref{fig:Max}, is very small.

\nocite{*}
\bibliography{RefOptimal2.bib}% Produces the bibliography via BibTeX.